%% file: main.tex
\documentclass[conference]{IEEEtran}
\IEEEoverridecommandlockouts

\input{preamble}

\input{header}

\begin{document}

\maketitle

\input{sections/abstract}

\input{sections/introduction}

\input{sections/system}

\input{sections/results}

\input{sections/conclusion}

\bibliographystyle{IEEEtran}
\bibliography{references}

\end{document}

%% file: preamble.tex
\usepackage{amsmath}
\usepackage{amssymb}
\usepackage{graphicx}
\usepackage{booktabs}
\usepackage{siunitx}
\usepackage{subcaption}
\usepackage{hyperref}
\usepackage{cleveref}
\usepackage{tikz}
\usepackage{pgfplots}
\pgfplotsset{compat=newest}
\usetikzlibrary{arrows.meta, positioning, shapes.geometric}

%% file: header.tex
\title{Real-Time Text Transmission via LLM-Based Entropy Coding over Fixed-Rate Channels}

\author{
  \IEEEauthorblockN{Vishnu Teja Kunde, Jean-Francois Chamberland, Krishna R. Narayanan}
  \IEEEauthorblockA{Texas A\&M University\\\texttt{\{vishnukunde,chmbrlnd,krn\}@tamu.edu}}
  \thanks{
    This material is based upon work supported, in part, by the National Science Foundation under Grant No.~CNS-2148354.}
  \and
  \IEEEauthorblockN{Jamison Ebert}
  \IEEEauthorblockA{Brigham Young University\\\texttt{jrebert@byu.edu}}
}

%% file: sections/abstract.tex
\begin{abstract}
Learning, prediction, and compression are intimately connected: a model that accurately predicts the next symbol in a sequence can be coupled with a source coder to compress that sequence near its information-theoretic limit.
When tokenized characters arriving at a fixed reading pace are encoded into variable-length codewords and streamed over a fixed-rate channel, a queue forms whose per-token delay depends on the mean and variance of the bit lengths and on the coder's algorithmic latency.
This paper investigates the compression--delay tradeoff that arises when a causal language model serves as the sequential predictor within a predict-then-code architecture for real-time text transmission.
Several coding schemes are compared: Shannon (ideal), Huffman, arithmetic coding, rANS at various block sizes, and gzip.
The analysis separates algorithmic delay, inherent to the coder, from computational delay, which shrinks as hardware improves.
Huffman is the practical choice for over-provisioned channels, with zero algorithmic delay and modest compression overhead.
Arithmetic coding achieves near-optimal compression at the cost of decodability delay.
Findings are validated across two scales: GPT-2 (124M) and Llama~3.2 (3B), a twenty-five-fold parameter range.
This scaling yields an approximately 38\% reduction in bits per character, effectively over-provisioning the channel and thereby changing which coder is optimal.
\end{abstract}

%% file: sections/introduction.tex
\section{Introduction}
\label{sec:intro}

Large language models (LLMs) have advanced rapidly in recent years, progressing from modest-scale architectures to billion-parameter transformers that capture long-range statistical dependencies in sequential data~\cite{radford2019language,meta2024llama3}.
A classical insight from information theory is that prediction and compression are two sides of the same coin: a model that assigns accurate conditional probabilities to the next symbol can be turned into an efficient compressor, because likely continuations require fewer bits to describe~\cite{shannon1951prediction,rissanen1984universal}.
The predict-then-code architecture formalizes this connection by pairing a sequential predictor with an entropy coder~\cite{cleary1984data}, and modern LLMs make it concrete by providing sharper conditional distributions over longer contexts than previously possible~\cite{deletang2024language,valmeekam2023llmzip}.
As these models grow in capability, the achievable compression rate drops, raising a natural question: how do the delay characteristics of different source coders interact with this improving source model?

Speech and language offer an ideal setting in which to study this question.
Unlike bulk data that can be compressed offline, spoken language is inherently a streaming source, produced and consumed in real time.
A live broadcast, a phone call, or a voice assistant all generate tokens at the pace of human speech, and the encoded bitstream must be delivered with low latency for the exchange to remain intelligible.
As an accessible proxy for such real-time sources, we consider text read at approximately 300~words per minute.
The character stream is passed through a tokenizer, and an LLM predictor supplies the conditional distributions used to entropy-code the tokens.
The resulting bit stream enters a queue served by a fixed-rate channel of capacity~$C$~bits per second, and each token's bit cost depends on how predictable it is given the preceding context.

The information content of token~$x_n$ given context~$\mathbf{x}_{<n}$ is
\begin{equation}
  b(x_n) = -\log_2 p(x_n | \mathbf{x}_{<n}),
  \label{eq:shannon}
\end{equation}
where $p$ is the true conditional distribution of the source.
Any lossless code may assign a particular token fewer or more bits than $b(x_n)$, but the entropy $H(X) = \mathbb{E}[b(X)]$ is the minimum achievable \emph{average} description length, a threshold known as the \emph{Shannon limit}: deviating from the information content for one token necessarily inflates the cost of others, leaving the expected rate unchanged or higher.
In practice, a model~$q$ approximates~$p$, and the average coding cost under~$q$ decomposes as
\begin{equation}
  H(p,q) = H(p) + D_{\mathrm{KL}}(p \| q),
  \label{eq:crossentropy}
\end{equation}
where $H(p)$ is the source entropy rate and the Kullback--Leibler (KL) divergence $D_{\mathrm{KL}}(p \| q)$ is the penalty for model mismatch.
Minimizing cross-entropy is, in fact, the predominant training objective for modern language models; every reduction in training loss corresponds directly to a reduction in the number of bits needed to describe the source.
This perspective connects compression to the minimum description length principle~\cite{barron1998minimum,rissanen1984universal}, and the rapid progress in LLM capability translates directly into tighter compression, lowering average bit cost and thus the required channel capacity.

Achieving~\eqref{eq:shannon} in practice requires an entropy coder that maps token probabilities to bit sequences, and the choice of coder introduces its own tradeoffs.
Arithmetic coding (AC)~\cite{rissanen1979,witten1987arithmetic} approaches the Shannon limit, but its streaming decoder must accumulate bits produced by future tokens before it can commit to the current one, introducing an \emph{intrinsic algorithmic delay} that persists regardless of implementation speed.
Huffman coding~\cite{huffman1952} avoids this delay entirely because each codeword is self-decodable, though it pays a price in integer-bit overhead.
Asymmetric numeral systems (rANS)~\cite{duda2009} with block size~$K$ occupy a tunable middle ground, where larger blocks improve compression at the cost of proportionally higher latency.
Gzip (DEFLATE~\cite{deutsch1996deflate} with \texttt{Z\_SYNC\_FLUSH} per token) offers immediate decodability but sacrifices cross-token compression, inflating bit costs by $10$--$12\times$ over Shannon.

While the predict-then-code literature has focused primarily on compression \emph{rate}, the \emph{delay} implications of different coding schemes under real-time constraints remain largely unexplored.
It is important to distinguish two sources of delay in such systems.
\emph{Computational delay} arises from the time needed to run the language model and the coder; it shrinks as hardware improves and, following the Bitter Lesson~\cite{sutton2019bitter}, can be expected to diminish over time.
\emph{Algorithmic delay}, by contrast, is inherent in the coder's structure: it reflects buffering, deferred decodability, or block formation, and it persists regardless of how fast the underlying computation becomes.
This article focuses on the latter as the irreducible bottleneck for real-time predict-then-code systems.

The contributions of this work are as follows.
\begin{enumerate}
  \item A queueing model for real-time text transmission is developed, in which variable-length codewords produced by a predict-then-code system are streamed over a constant-rate channel.
  Within this model, Shannon (ideal), Huffman, arithmetic coding, rANS at several block sizes, and gzip are compared in terms of both compression efficiency and algorithmic delay.
  \item The impact of language model quality on the compression--delay tradeoff is assessed empirically by running identical experiments under GPT-2 (124M parameters) and Llama~3.2 (3B parameters), spanning a $25\times$ range in model scale.
  \item The compression--latency Pareto frontier across coding schemes and block sizes is identified, revealing operating regimes in which different coders are preferred.
\end{enumerate}

%% file: sections/system.tex
\section{System Model}
\label{sec:system}

\begin{figure}[t]
  \centering
  \input{figures/system_model}
  \caption{Notional diagram for streaming \emph{predict-then-code} transmission.
  Characters arrive at rate~$\lambda$ characters per second (cps), are encoded using the LLM's conditional distributions, queued, and served through a constant-rate channel.
  The delay $D(n)$ for token~$n$ is the elapsed time between the moment the token is generated at the source and the moment it is successfully decoded at the destination.}
  \label{fig:system_model}
\end{figure}
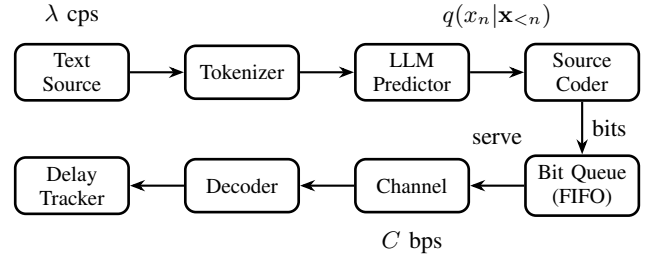

This section describes the predict-then-code pipeline used to study the compression--delay tradeoff in real-time text transmission.
A text source generates characters at a fixed rate, a tokenizer groups them into tokens, a causal language model assigns conditional probabilities to each token, and a source coder converts those probabilities into a variable-length bitstream.
The bits enter a FIFO queue served at constant rate by the channel, and a delay tracker records when each token becomes decodable at the receiver.
\Cref{fig:system_model} provides a high-level view of the system.

\subsection{Language Model Predictor}
\label{sec:lm_predictor}

A causal language model (LM) defines a conditional probability distribution $q(x_n | x_1, \ldots, x_{n-1}) = q(x_n | \mathbf{x}_{< n})$ over a vocabulary~$\mathcal{V}$ of size~$V$.
Given a sequence of tokens, the LM produces a softmax distribution over the next token, which the source coder uses to assign bit costs.
The quality of this distribution directly determines compression efficiency via~\eqref{eq:crossentropy}.
A model that assigns conditional probabilities closer to the true distribution $p(x_n | \mathbf{x}_{< n})$ yields fewer coded bits on average.
Since modern LMs are trained by minimizing cross-entropy, reductions in training loss translate directly to shorter code lengths when coupled with an entropy coder~\cite{rissanen1984universal}.
The LM serves as a sequential probability engine whose quality governs compression performance~\cite{deletang2024language}.

Two transformer-based~\cite{vaswani2017attention} causal LLMs are considered:
\begin{itemize}
  \item \textbf{GPT-2}~\cite{radford2019language}: 124M parameters, $V = 50{,}257$, 12-layer transformer decoder.
  \item \textbf{Llama~3.2~3B}~\cite{meta2024llama3}: 3.2B parameters, $V = 128{,}256$, with grouped-query attention and RoPE embeddings.
\end{itemize}
Llama~3.2 is roughly twenty-five times larger than GPT-2 in parameter count.

\subsection{Source Coding Schemes}
\label{sec:coders}

Six coding schemes spanning the compression--latency spectrum are evaluated.
All LLM-based schemes instantiate the classical predict-then-code architecture~\cite{cleary1984data,rissanen1984universal}: the LLM supplies sequential probability assignments $q(x_n | \mathbf{x}_{<n})$ and the coder converts them to bits.
Only gzip departs from this paradigm, relying on its own internal dictionary model instead of the LLM's predictions.

\subsubsection{Shannon Coding (Approximate Lower Bound)}
\label{sec:shannon}

The Shannon coder assigns each token a code length equal to its log-loss under the model:
\begin{equation}
  b_S(x_n) = -\log_2 q(x_n | \mathbf{x}_{<n}).
  \label{eq:shannon_bits}
\end{equation}
This can produce fractional bits and is not realizable in practice, but serves as a benchmark: an approximate lower bound on the code length achievable by any token-by-token coder operating under model~$q$.
When applied to the true conditional distribution $p(x_n | \mathbf{x}_{<n})$, the Shannon coder achieves the entropy rate of the source and is therefore an actual lower bound.
Applied instead to the postulated distribution $q(x_n | \mathbf{x}_{<n})$, it incurs a mismatch penalty akin to \eqref{eq:crossentropy}, so this benchmark becomes increasingly tight as $q$ approaches $p$.
The Shannon coder has zero algorithmic delay: each token's bit cost is self-contained.

\subsubsection{Huffman Coding}
\label{sec:huffman}

Huffman coding~\cite{huffman1952} assigns integer-length prefix-free codewords.
Unlike the classical setting of a static source PMF, the Huffman code here is constructed anew at each token position from the LLM's conditional distribution $q(x_n | \mathbf{x}_{<n})$, making it an adaptive, context-dependent code.
The resulting codeword length satisfies
\begin{equation}
  b_H(x_n) = \max\big(1, \lceil -\log_2 q(x_n | \mathbf{x}_{<n}) \rceil\big).
  \label{eq:huffman_bits}
\end{equation}
The $\lceil \cdot \rceil$ rounding and the minimum of 1~bit per token introduce overhead of at most 1~bit per symbol.
In practice, across the corpus used here, Huffman averages ${\sim}9\%$ overhead relative to Shannon.
Like Shannon coding, Huffman has zero algorithmic delay: each codeword is immediately decodable.

\subsubsection{Arithmetic Coding}
\label{sec:ac}

Arithmetic coding~\cite{rissanen1979,witten1987arithmetic} represents an entire sequence as a single number in $[0, 1)$, with the interval progressively narrowed by each token's probability.
It is the canonical realization of predict-then-code: unlike Huffman, AC imposes no integer-length constraint, allowing code lengths to track the predictor's log-loss with high fidelity.
A streaming implementation with 64-bit precision emits bits as the interval narrows sufficiently, achieving near-optimal compression, within fractions of a bit of Shannon.

However, the streaming AC decoder introduces \emph{deferred decodability}.
The decoder must accumulate enough bits in its state register before it can commit to a token.
These bits may originate from \emph{future} tokens that have not yet arrived at the encoder.
The decodability time for token~$n$ is
\begin{equation}
  D_{\mathrm{AC}}(n) = \max\big(t_{\mathrm{serve}}(\beta_n), t_{\mathrm{arr}}(n)\big) - t_{\mathrm{arr}}(n),
  \label{eq:ac_delay}
\end{equation}
where $\beta_n$ is the index of the last bit the decoder must consume to resolve token~$n$, and $t_{\mathrm{serve}}(\beta_n)$ is the time that bit exits the channel.
With 64-bit precision, the decoder requires roughly 64~bits of initialization before any token can be decoded.
At approximately 5~bits per token, this translates to about 13~tokens of lookahead, or just over 2.5~seconds at the source character rate of $\lambda = 20$~characters per second.

\subsubsection{rANS with Block Size $K$}
\label{sec:rans}

Asymmetric numeral systems (ANS)~\cite{duda2009} encode information into a single integer state using a LIFO (last-in, first-out) structure.
In the rANS variant with block size~$K$, the encoder buffers $K$~tokens, encodes them in reverse order into a state integer, then emits the resulting bits.
The decoder reconstructs tokens by reversing this process.

Because the entire block must be buffered before encoding begins, the first token in each block is held until the remaining $K-1$ tokens of the block arrive.
Under the constant-character-rate source considered here, token inter-arrival times are proportional to the per-token character counts $B_n$ and therefore vary across the stream, so the buffering delay is itself a random variable.
In expectation, the first-token delay satisfies
\begin{equation}
  \mathbb{E}[D_{\mathrm{rANS}}(n)] \geq (K-1) \frac{\mathbb{E}[B]}{\lambda},
  \label{eq:rans_delay}
\end{equation}
where $\lambda$ is the source character rate and $\mathbb{E}[B]$ is the tokenizer's mean characters per token.
For example, at $\lambda = 20$~characters per second with $\mathbb{E}[B] \approx 4$ and $K = 16$, the expected delay is roughly 3~seconds for the first token of each block.

Larger~$K$ improves compression by amortizing per-block state overhead across more symbols.
The per-block overhead consists of $\lceil \log_2(\text{state range}) \rceil$~bits to finalize the encoder state, where the state range is an rANS implementation parameter.
For instance, a 32-bit rANS implementation pays 32~bits per block, while a 64-bit implementation pays 64~bits.
This overhead is divided among $K$~symbols, so $K = 1$ pays the full 32-bit penalty per token while $K = 16$ pays only 2~bits per token.

\subsubsection{Gzip (DEFLATE with Per-Token Flush)}
\label{sec:gzip}

The gzip coder uses zlib's DEFLATE algorithm~\cite{deutsch1996deflate} with \texttt{Z\_SYNC\_FLUSH} after every token.
This forces the encoder to emit a complete, self-decodable block at each token boundary, enabling immediate decodability so that the decoder never waits for future data.

The cost is severe: flushing prevents cross-token compression and adds block framing overhead.
Gzip does not use the LLM's probability distribution at all; it relies on its own LZ77 dictionary, which is poorly suited to single-token inputs.
The result is ten to twelve times more bits per token than Shannon coding.

\subsection{Channel and Bit Queue}
\label{sec:channel}

The average source rate in bits per second (bps) is
\begin{equation}
  R_s = \lambda \cdot \bar{\ell}_s,
  \label{eq:source_rate}
\end{equation}
where $\lambda$ is the source character rate and $\bar{\ell}_s$ is the mean code length in bits per character under coding scheme~$s$.
Reporting the code length in bits per character keeps the source rate $R_s$ tokenizer-independent: the same physical reading speed $\lambda$ applies regardless of how the tokenizer segments the stream.
The channel serves bits in first-in, first-out (FIFO) order at rate~$C$~bps.
For scalar coders (Shannon, Huffman, gzip), each token's bits are encoded separately and the per-token delay satisfies Lindley's recursion
\begin{equation}
  D(n) = \max\big(0, t_{\mathrm{exit}}(n-1) - t_{\mathrm{arr}}(n)\big) + \frac{b(x_n)}{C},
  \label{eq:lindley}
\end{equation}
with $t_{\mathrm{exit}}(0) = 0$.
The first term is the waiting time, positive only when token~$n$ arrives while the channel is still serving token~$n-1$; the second term $b(x_n)/C$ is the service time.
$D(n)$ measures the elapsed time between token~$n$'s generation at the source and its successful decoding at the destination, and $t_{\mathrm{exit}}(n) = t_{\mathrm{arr}}(n) + D(n)$ propagates the recursion.
When the channel rate exceeds the source rate ($C > R_s$), the queue is stable and delays are bounded.
When $C < R_s$, the queue grows without bound and delays diverge.
For AC, the decodability delay $D_{\mathrm{AC}}(n)$ in~\eqref{eq:ac_delay} accounts for deferred decoding.
For rANS, $D_{\mathrm{rANS}}(n)$ is subject to a similar mechanisms where completion time depends, partly, on the location of the token within the block; it is bounded below by~\eqref{eq:rans_delay} due to block buffering.

For arithmetic coding, a specialized adapter tracks per-bit service times through the channel.
The decoder's \texttt{bits\_consumed} counter determines which channel bit must have exited before a token becomes decodable.
This captures the deferred-decodability delay that distinguishes AC from scalar coders.

For rANS, the encoder buffers all $K$ tokens in a block before encoding them jointly.
The resulting bits are emitted and queued as a single unit, so no token in the block can be decoded until the last bit of that block exits the channel.
The tokens within a block are then recovered in reverse order.
As such, the first token in the block must wait for both the buffering of $K-1$ subsequent tokens and the channel service of the entire block.

The simulation uses a fully deterministic synthetic clock in which token~$i$ arrives at the encoder when its last character has been emitted by the source: $t_{\mathrm{arr}}(i) = \big(\sum_{j \leq i} B_j\big) / \lambda$, where $B_j$ is the number of characters in token~$j$.
There is no wall-clock timing or threading, ensuring reproducible results and isolating algorithmic delay from implementation artifacts.

%% file: figures/system_model.tex
\begin{tikzpicture}[
  font=\small, line width=1pt, node distance=0.55cm and 0.7cm,
  block/.style={draw, rounded corners, minimum height=0.7cm, minimum width=1.5cm, font=\footnotesize, align=center},
  arr/.style={-{Stealth[length=2mm]}, thick}
]
  \node[block] (src) at (0,0.75) {Text\\Source};
  \node[block] (tok) at (2.25,0.75) {Tokenizer};
  \node[block] (lm) at (4.5,0.75) {LLM\\Predictor};
  \node[block] (enc) at (6.75,0.75) {Source\\Coder};
  \node[block] (queue) at (6.75,-0.75) {Bit Queue\\(FIFO)};
  \node[block] (chan) at (4.5,-0.75) {Channel};
  \node[block] (dec) at (2.25,-0.75) {Decoder};
  \node[block] (track)  at (0,-0.75) {Delay\\Tracker};

  \draw[arr] (src) -- (tok);
  \draw[arr] (tok) -- (lm);
  \draw[arr] (lm) -- (enc);
  \draw[arr] (enc) -- (queue) node[midway, right] {bits};
  \draw[arr] (queue) -- (chan);
  \draw[arr] (chan) -- (dec);
  \draw[arr] (dec) -- (track);

  \node at (0,1.5) {$\lambda$ cps};
  \node at (5.625,1.5) {$q(x_n|\mathbf{x}_{<n})$};
  \node at (5.625,-0.125) {serve};
  \node at (4.5,-1.5){$C$ bps};
\end{tikzpicture}

%% file: sections/results.tex
\section{Results}
\label{sec:results}

Five public-domain texts from Project Gutenberg~\cite{gutenberg} serve as test sources: \emph{Frankenstein} (Shelley), \emph{Leaves of Grass} (Whitman), \emph{Moby-Dick} (Melville), \emph{Pride and Prejudice} (Austen), and \emph{The Republic} (Plato/Jowett).
Each text is tokenized using the HuggingFace tokenizer~\cite{wolf2020transformers} corresponding to the language model (GPT-2 BPE~\cite{sennrich2016neural} or Llama SentencePiece).
Because the two tokenizers produce tokens of different granularity, cross-model comparisons are reported in bits per character (bpc), obtained by dividing bits per token by the mean characters per token $\mathbb{E}[B]$ for the corresponding tokenizer.
Across the five texts, GPT-2 averages $\mathbb{E}[B] \approx 3.84$ characters per token while Llama averages $\mathbb{E}[B] \approx 4.17$.

Both implementations are driven by the same physical source: a character stream at $\lambda = 20$~characters per second (cps), corresponding to roughly 240~words per minute.
The character rate is the fundamental quantity here because it is tokenizer-independent; the resulting token arrival rate $\lambda / \mathbb{E}[B]$ is text-dependent and ranges from 4.85 to 5.78~tokens/s for GPT-2 and from 4.67 to 5.04~tokens/s for Llama.
The channel rate is swept as $C = \alpha \cdot \lambda \cdot \bar{\ell}_{\mathrm{Sh}}$, where $\bar{\ell}_{\mathrm{Sh}}$ is the Shannon mean code length in bits per character under the LLM and $\alpha \in \{0.8, 0.9, 0.95, 0.98, 1, 1.02, 1.05, 1.1, 1.2, 1.5, 2\}$.
Because $\bar{\ell}_{\mathrm{Sh}}$ is the cross-entropy of the source under the LLM (the Shannon lower bound on token-by-token coding under model~$q$ and an approximation to the true Shannon rate that tightens as $q \to p$, cf.~\eqref{eq:crossentropy}), the ratio $\alpha = 1.0$ corresponds to a channel rate equal to the model-based source rate $\lambda \bar{\ell}_{\mathrm{Sh}}$; values below unity push the system into overload, while values above unity provide increasing headroom.
All experiments use 10{,}000 tokens per (text, model) pair; under a fixed character rate, this corresponds to roughly 33--42 thousand characters of source text depending on the tokenizer.

For each coding scheme and channel rate ratio~$\alpha$, two metrics are reported: the mean code length in bits per character (and its overhead relative to Shannon), and the mean and 95th-percentile per-token delay $D(n)$.

\subsection{Cross-Entropy by Model}
\label{sec:crossentropy_results}

\Cref{tab:crossentropy} compares the cross-entropy of GPT-2 and Llama~3.2 on the five Project Gutenberg texts, each evaluated over 10{,}000~tokens.
Both bits per token (bpt) and bits per character (bpc) are reported; the bpc column normalizes for tokenizer granularity and offers an appropriate basis for cross-model comparison.

\begin{table}[t]
  \centering
  \caption{Cross-entropy at 10K tokens, reported in bits per token (bpt) and bits per character (bpc). The reduction column is computed on a bpc basis.}
  \label{tab:crossentropy}
  \sisetup{round-mode=places, round-precision=2}
  \begin{tabular}{l S[table-format=1.2] S[table-format=1.2] S[table-format=1.2] S[table-format=1.2] S[table-format=2.0, round-precision=0]}
    \toprule
    & \multicolumn{2}{c}{GPT-2} & \multicolumn{2}{c}{Llama 3.2} & \\
    \cmidrule(lr){2-3} \cmidrule(lr){4-5}
    {Text} & {bpt} & {bpc} & {bpt} & {bpc} & {Reduction (\%)} \\
    \midrule
    Frankenstein       & 5.74 & 1.39 & 3.49 & 0.81 & 42 \\
    Leaves of Grass    & 6.16 & 1.78 & 4.58 & 1.13 & 37 \\
    Moby-Dick          & 6.11 & 1.64 & 3.54 & 0.89 & 46 \\
    Pride \& Prejudice & 4.47 & 1.17 & 3.54 & 0.83 & 29 \\
    The Republic       & 5.52 & 1.35 & 3.71 & 0.87 & 36 \\
    \midrule
    \textbf{Mean}      & \textbf{5.60} & \textbf{1.47} & \textbf{3.77} & \textbf{0.91} & \textbf{38} \\
    \bottomrule
  \end{tabular}
\end{table}

On a bpc basis, Llama~3.2 reduces cross-entropy by 38\% relative to GPT-2, a substantial gain achieved at roughly twenty-five times the computational cost.

We emphasize that the reduction varies by text: highly predictable prose (Pride~\&~Prejudice) sees a 29\% improvement in bpc, while Moby-Dick, with its archaic and varied vocabulary, benefits by 46\%.
This confirms that the KL divergence term in~\eqref{eq:crossentropy} is text-dependent and that model scaling yields diminishing returns for already-predictable sources.

\subsection{Compression Efficiency}
\label{sec:compression}

\Cref{tab:bits} reports the mean bits per character for each coding scheme under both GPT-2 and Llama~3.2, averaged across all five texts at 10{,}000~tokens.
AC achieves essentially zero overhead under both models, confirming that 64-bit streaming arithmetic coding is effectively optimal for individually coded tokens.
Huffman achieves near-zero overhead under GPT-2 and approximately 5\% under Llama, well within the theoretical bound of at most 1~bit per symbol.

\begin{table}[t]
  \centering
  \caption{Mean bits per character (bpc) by coder at 10K tokens, averaged across 5~texts. Overhead is relative to Shannon under the same LLM.}
  \label{tab:bits}
  \begin{tabular}{l S[table-format=2.2] S[table-format=4.0, round-precision=0] S[table-format=2.2] S[table-format=4.0, round-precision=0]}
    \toprule
    & \multicolumn{2}{c}{GPT-2} & \multicolumn{2}{c}{Llama 3.2} \\
    \cmidrule(lr){2-3} \cmidrule(lr){4-5}
    {Coder} & {bpc} & {Overhead (\%)} & {bpc} & {Overhead (\%)} \\
    \midrule
    Shannon   &  1.47 &  {---} &  0.91 &  {---} \\
    AC        &  1.47 &   0    &  0.91 &   0 \\
    Huffman   &  1.45 &  -1    &  0.96 &   5 \\
    rANS-K16  &  1.99 &  35    &  1.40 &  54 \\
    rANS-K8   &  2.53 &  72    &  1.88 & 107 \\
    rANS-K4   &  3.59 & 144    &  2.86 & 214 \\
    rANS-K1   &  9.98 & 579    &  8.74 & 860 \\
    Gzip      & 17.74 & {1107} & 16.74 & {1740} \\
    \bottomrule
  \end{tabular}
\end{table}

The rANS results reveal the per-block state overhead.
At $K = 1$, each token pays the full 32-bit state finalization cost, producing roughly 10~bpc under GPT-2, nearly $7\times$~Shannon.
As $K$ increases, the overhead amortizes: $K = 16$ achieves 1.99~bpc under GPT-2 (35\% overhead), and extrapolation suggests $K \geq 64$ would approach AC-level efficiency.

Gzip at approximately 18~bpc under GPT-2 ($12\times$~Shannon) highlights a structural mismatch with the real-time streaming setting.
The per-token \texttt{Z\_SYNC\_FLUSH} is not an arbitrary handicap: in a delay-sensitive application, the coder must commit bits to the channel promptly, and allowing gzip to accumulate a longer window would trade compression for buffering delay, the same tradeoff studied throughout this paper.
Flushing per token is the natural choice for immediate decodability, just as Huffman emits a self-decodable codeword per token.
The difference is that gzip cannot exploit the shared LLM predictor; it must discover statistical structure on its own from the raw byte stream, and single-token inputs give its LZ77 dictionary almost nothing to work with.

\begin{figure}[t]
  \centering
  \begin{subfigure}[t]{\columnwidth}
    \centering
    \input{figures/bpc_comparison_gpt2}
    \caption{GPT-2 (124M parameters)}
    \label{fig:bits_gpt2}
  \end{subfigure}
  \\[0.5em]
  \begin{subfigure}[t]{\columnwidth}
    \centering
    \input{figures/bpc_comparison_llama}
    \caption{Llama 3.2 (3B parameters)}
    \label{fig:bits_llama}
  \end{subfigure}
  \caption{Bits per character by coding scheme and text. Both plots share the same log-scale vertical axis for direct comparison. Shannon and AC overlap in both models.}
  \label{fig:bits}
\end{figure}
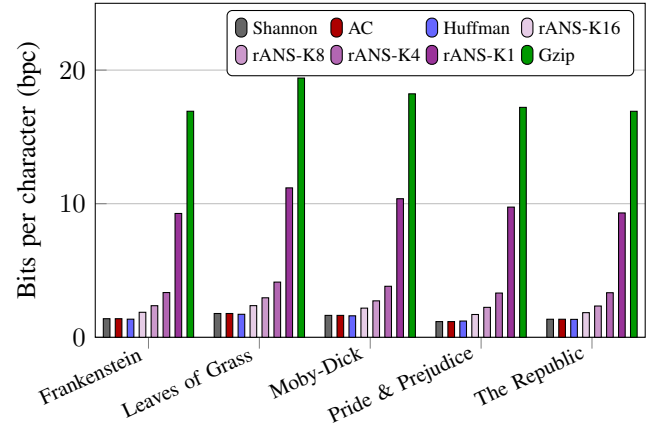
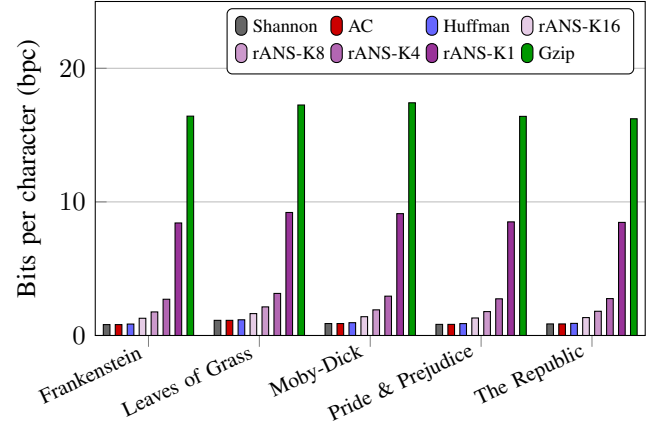

This contrast underscores why the predict-then-code coders perform well at the token granularity.
Modern tokenizers (BPE, SentencePiece) already group characters into semantically meaningful chunks of three to four characters on average, providing a coding unit that is large enough to amortize overhead yet small enough to maintain low delay.
The LLM then supplies a sharp conditional distribution over these tokens.
Together, tokenization and prediction make token-by-token coding viable in a way that character-by-character coding with a dictionary compressor is not.
\Cref{fig:bits} shows the per-text breakdown for both GPT-2 and Llama~3.2.

\subsection{Delay vs.\ Channel Rate}
\label{sec:delay}

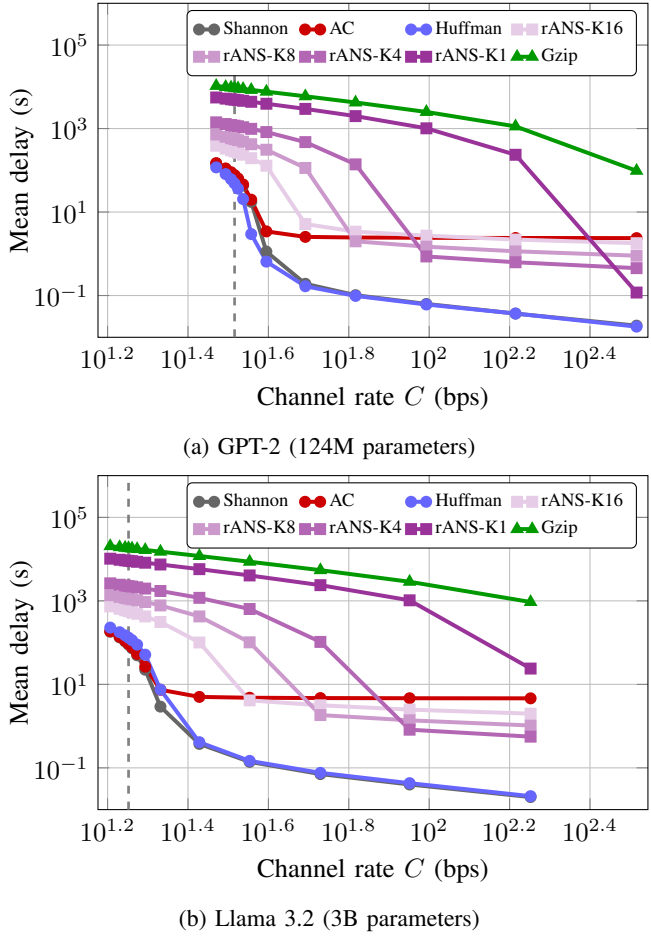
\begin{figure}[t]
  \centering
  \begin{subfigure}[t]{\columnwidth}
    \centering
    \input{figures/delay_vs_rate_moby_dick_gpt2}
    \caption{GPT-2 (124M parameters)}
    \label{fig:delay_gpt2}
  \end{subfigure}
  \\[0.5em]
  \begin{subfigure}[t]{\columnwidth}
    \centering
    \input{figures/delay_vs_rate_moby_dick_llama}
    \caption{Llama 3.2 (3B parameters)}
    \label{fig:delay_llama}
  \end{subfigure}
  \caption{Mean delay vs.\ channel rate $C$ in bps (Moby-Dick) averaged over 10K tokens.
  The vertical dashed line in each plot marks the model-based source rate $\lambda \bar{\ell}_{\mathrm{Sh}}$ for the corresponding LLM (32.8~bps for GPT-2, 17.9~bps for Llama).
  Both plots share the same axes, making the effect of model quality directly visible.}
  \label{fig:delay}
\end{figure}

\Cref{fig:delay} shows mean delay versus absolute channel rate $C$ (in bps) for \emph{Moby-Dick} at 10{,}000~tokens under both GPT-2 and Llama~3.2.
Both plots share the same axis ranges for direct comparison; the vertical dashed line in each marks the model-based source rate $\lambda \bar{\ell}_{\mathrm{Sh}}$ ($\alpha = 1.0$) for the corresponding LLM.
Because the physical reading speed is matched across models, the same value of $C$ represents the same physical channel in both plots, making Llama's lower capacity requirement immediately visible.
We can characterize performance in various regimes.

\textbf{Over-provisioned (rate $\gg$ Shannon):} The delays of the Shannon lower bound and Huffman coder drop below 1~second.
The queue rarely builds up, and each token's bits are served almost immediately.
Huffman's small overhead is inconsequential because the channel has ample spare capacity.
The delays of other schemes are mainly algorithmic; transmission delays are less significant.
In this regime, Huffman is the practical choice: zero algorithmic delay, simple implementation, and adequate compression.

\textbf{Near capacity (rate $<$ Shannon):} Delays grow sharply as the queue approaches instability.
Under model match ($q \approx p$), Huffman's small positive overhead implies that it reaches instability slightly sooner than the Shannon coder, while AC, despite its decodability delay, maintains competitive mean delay because its near-optimal compression keeps the queue stable.
Model friction refines this picture: Huffman's mean code length is bounded below by the true source entropy $H(p)$, not by the model cross-entropy $H(p, q) = \bar{\ell}_{\mathrm{Sh}}$, so favorable integer rounding of the codeword lengths can leave Huffman below the cross-entropy on a particular sample whenever the KL gap $D_{\mathrm{KL}}(p \| q)$ is large enough to absorb the rounding.
This is what \Cref{tab:bits} shows under GPT-2, where Huffman achieves $-1\%$ overhead and is therefore marginally more stable than the Shannon coder over a narrow range of $\alpha$.

\textbf{Below capacity (rate $\ll$ Shannon):} In this regime, all pragmatic coders are unstable with unbounded delay growth.
Gzip requires sizable channel rates to avoid queue instability and is effectively unusable in bandwidth-constrained scenarios.

\subsection{rANS Block Size Tradeoff}
\label{sec:rans_tradeoff}

\Cref{fig:delay} also resolves the rANS family across block sizes $K \in \{1, 2, 4, 8, 16\}$ for \emph{Moby-Dick} under Llama~3.2, with the other schemes serving as reference.
The expected floor delay follows directly from~\eqref{eq:rans_delay}: $\mathbb{E}[D_{\min}] = (K-1) \mathbb{E}[B] / \lambda$.
At $K = 1$, rANS has zero buffering delay but 36.2~bits/token (9.12~bpc); the per-symbol state overhead dominates.
At $K = 16$, compression improves to 5.6~bits/token (1.41~bpc, 57\% overhead over Shannon), but the first token in each block waits on average about 3~seconds.
The per-K curves in \Cref{fig:delay} expose the resulting compression--latency Pareto frontier: no block size simultaneously minimizes both compression overhead and latency, and the lower envelope across $K$ approaches Shannon's delay profile in the well-provisioned regime, where at each channel rate some $K$ optimally balances overhead reduction against buffering delay.

However, modern tokenizers substantially reduce the practical value of this tradeoff.
BPE and SentencePiece already group characters into semantically meaningful units of three to four characters on average, providing a natural coding granularity.
In the classical character-level or bit-level setting, Huffman's worst-case overhead of 1~bit per symbol is large relative to a per-character entropy near 1~bpc; grouping via rANS was essential to amortize this cost.
At the token level, the same 1-bit overhead is amortized over a conditional entropy of several bits per token (e.g., 3.5~bits per token, equivalently 0.89~bpc, for Llama on Moby-Dick), reducing the worst-case penalty to roughly 29\% and the observed penalty to near zero under GPT-2 and approximately 5\% under Llama (\Cref{tab:bits}).
Combined with a strong LLM predictor that sharpens the conditional distribution, Huffman achieves near-Shannon compression with zero algorithmic delay, substantially narrowing the operating regime where rANS's compression advantage justifies its buffering cost.

\subsection{Model Quality Impact}
\label{sec:model_quality}

Llama~3.2's 38\% reduction in bits per character over GPT-2 (\Cref{tab:crossentropy}) has a direct impact on delay through the source rate.
For a given channel rate~$C$, lower bpc means $R_s = \lambda \cdot \bar{\ell}_s$ decreases, effectively increasing the operating ratio~$\alpha$.
At a fixed channel rate of \SI{30}{bps} (a typical GPT-2 model-based source rate under the matched character rate), GPT-2 operates near $\alpha = 1.0$ on most texts, while Llama operates at $\alpha \approx 1.5$--$1.7$.
This shift from near-capacity to over-provisioned operation dramatically reduces queueing delay, from seconds to milliseconds for scalar coders.

The implication is that model quality and coder choice interact: with a weak model, near-capacity operation demands AC for its compression efficiency; with a strong model, the channel is effectively over-provisioned and simpler coders suffice.
Spending endpoint compute to sharpen the predictor and reduce source rate is a modern instantiation of the predict-then-code tradeoff~\cite{rissanen1984universal}: when the LLM is already deployed on-device, the marginal inference cost yields disproportionate delay reduction near the capacity boundary.
In our experiments, the roughly twenty-five-fold increase in model parameters offers a 38\% bpc gain, which exhibits diminishing returns in isolation but whose impact is amplified near the capacity boundary where small reductions in source rate result in large delay improvements.

\subsection{Practical Implications for Speech}
\label{sec:practical}

The preceding results assume a fixed character rate matched to a nominal reading speed.
In practice, human speakers exhibit substantial variability in rate, and a communication system must provision the channel for the fastest speakers it is designed to serve.
This approach inherently over-provisions the channel for speakers of average or below-average pace, pushing the system into the regime where queueing delay is low for all coders.
In such a regime, the results of this study point to a clear practical recommendation: Huffman coding paired with an LLM predictor is difficult to improve upon.
It offers zero algorithmic delay, simple implementation, and compression overhead that is modest relative to the token-level conditional entropy.
Neither arithmetic coding's deferred decodability nor rANS's block buffering provides a meaningful advantage when the channel already has headroom.

This observation shifts the fundamental design tradeoff away from the choice of coder and toward the choice of predictor.
Llama~3.2 offers a significant performance improvement over GPT-2, but at the cost of substantially greater model complexity and inference latency.
The Bitter Lesson~\cite{sutton2019bitter} suggests planning for increasingly powerful compute, which would favor deploying the largest feasible model.
At the same time, there is a practical middle ground: lightweight models obtained through distillation, quantization, or domain-specific adaptation may achieve sharp conditional distributions at inference speeds compatible with real-time speech.
The design challenge, in this view, is no longer primarily algorithmic (which coder to use) but rather one of AI engineering: finding the right balance between predictive quality and inference cost for a given hardware and latency budget.

%% file: figures/bpc_comparison_gpt2.tex
\begin{tikzpicture}
\begin{axis}[
    ybar,
    legend image code/.code={
        \draw[#1] (0cm,-0.1cm) rectangle (0.15cm,0.15cm);
    },
    bar width=2.5pt,
    width=\columnwidth,
    height=6cm,
    ymin=0,
    ymax=25,
    ylabel={Bits per character (bpc)},
    symbolic x coords={Frankenstein, Leaves of Grass, Moby-Dick, Pride \& Prejudice, The Republic},
    xtick=data,
    x tick label style={rotate=25, anchor=east, font=\footnotesize},
    legend style={at={(0.98,0.98)}, anchor=north east, font=\scriptsize, legend columns=4, rounded corners=2pt},
    legend cell align={left},
    enlarge x limits=0.12,
    ymajorgrids=true,
    xtick pos=bottom,
    ytick pos=left,
]

\addplot[fill=black!60!white] coordinates {
    (Frankenstein, 1.3918) (Leaves of Grass, 1.7809) (Moby-Dick, 1.6398)
    (Pride \& Prejudice, 1.1747) (The Republic, 1.3515)
};

\addplot[fill=red!68!black] coordinates {
    (Frankenstein, 1.3918) (Leaves of Grass, 1.7810) (Moby-Dick, 1.6399)
    (Pride \& Prejudice, 1.1747) (The Republic, 1.3515)
};

\addplot[fill=blue!60!white] coordinates {
    (Frankenstein, 1.3564) (Leaves of Grass, 1.7220) (Moby-Dick, 1.6082)
    (Pride \& Prejudice, 1.2163) (The Republic, 1.3424)
};

\addplot[fill=violet!20!white] coordinates {
    (Frankenstein, 1.8718) (Leaves of Grass, 2.3686) (Moby-Dick, 2.1827)
    (Pride \& Prejudice, 1.7065) (The Republic, 1.8443)
};

\addplot[fill=violet!40!white] coordinates {
    (Frankenstein, 2.3651) (Leaves of Grass, 2.9567) (Moby-Dick, 2.7276)
    (Pride \& Prejudice, 2.2413) (The Republic, 2.3413)
};

\addplot[fill=violet!60!white] coordinates {
    (Frankenstein, 3.3497) (Leaves of Grass, 4.1311) (Moby-Dick, 3.8187)
    (Pride \& Prejudice, 3.3117) (The Republic, 3.3362)
};

\addplot[fill=violet!80!white] coordinates {
    (Frankenstein, 9.2687) (Leaves of Grass, 11.1844) (Moby-Dick, 10.3727)
    (Pride \& Prejudice, 9.7444) (The Republic, 9.3079)
};

\addplot[fill=green!60!black] coordinates {
    (Frankenstein, 16.9202) (Leaves of Grass, 19.4043) (Moby-Dick, 18.2242)
    (Pride \& Prejudice, 17.2108) (The Republic, 16.9173)
};
\legend{Shannon, AC, Huffman, rANS-K16, rANS-K8, rANS-K4, rANS-K1, Gzip}
\end{axis}
\end{tikzpicture}

%% file: figures/bpc_comparison_llama.tex
\begin{tikzpicture}
\begin{axis}[
    ybar,
    legend image code/.code={
        \draw[#1] (0cm,-0.1cm) rectangle (0.15cm,0.15cm);
    },
    bar width=2.5pt,
    width=\columnwidth,
    height=6cm,
    ymin=0,
    ymax=25,
    ylabel={Bits per character (bpc)},
    symbolic x coords={Frankenstein, Leaves of Grass, Moby-Dick, Pride \& Prejudice, The Republic},
    xtick=data,
    x tick label style={rotate=25, anchor=east, font=\footnotesize},
    legend style={at={(0.98,0.98)}, anchor=north east, font=\scriptsize, legend columns=4, rounded corners=2pt},
    legend cell align={left},
    enlarge x limits=0.12,
    ymajorgrids=true,
    xtick pos=bottom,
    ytick pos=left,
]

\addplot[fill=black!60!white] coordinates {
    (Frankenstein, 0.8145) (Leaves of Grass, 1.1340) (Moby-Dick, 0.8937)
    (Pride \& Prejudice, 0.8321) (The Republic, 0.8652)
};

\addplot[fill=red!80!black] coordinates {
    (Frankenstein, 0.8145) (Leaves of Grass, 1.1341) (Moby-Dick, 0.8938)
    (Pride \& Prejudice, 0.8322) (The Republic, 0.8653)
};

\addplot[fill=blue!60!white] coordinates {
    (Frankenstein, 0.8556) (Leaves of Grass, 1.1755) (Moby-Dick, 0.9579)
    (Pride \& Prejudice, 0.8901) (The Republic, 0.9067)
};

\addplot[fill=violet!20!white] coordinates {
    (Frankenstein, 1.2893) (Leaves of Grass, 1.6382) (Moby-Dick, 1.4064)
    (Pride \& Prejudice, 1.3104) (The Republic, 1.3394)
};

\addplot[fill=violet!40!white] coordinates {
    (Frankenstein, 1.7638) (Leaves of Grass, 2.1411) (Moby-Dick, 1.9188)
    (Pride \& Prejudice, 1.7880) (The Republic, 1.8131)
};

\addplot[fill=violet!60!white] coordinates {
    (Frankenstein, 2.7129) (Leaves of Grass, 3.1482) (Moby-Dick, 2.9441)
    (Pride \& Prejudice, 2.7455) (The Republic, 2.7618)
};

\addplot[fill=violet!80!white] coordinates {
    (Frankenstein, 8.4226) (Leaves of Grass, 9.2077) (Moby-Dick, 9.1204)
    (Pride \& Prejudice, 8.5049) (The Republic, 8.4654)
};

\addplot[fill=green!60!black] coordinates {
    (Frankenstein, 16.4186) (Leaves of Grass, 17.2525) (Moby-Dick, 17.4185)
    (Pride \& Prejudice, 16.4019) (The Republic, 16.2225)
};
\legend{Shannon, AC, Huffman, rANS-K16, rANS-K8, rANS-K4, rANS-K1, Gzip}
\end{axis}
\end{tikzpicture}

%% file: figures/delay_vs_rate_moby_dick_gpt2.tex
\begin{tikzpicture}
\begin{axis}[
    width=\columnwidth,
    height=6cm,
    xmode=log,
    ymode=log,
    xmin=15,
    xmax=350,
    ymin=0.01,
    ymax=1000000,
    xlabel={Channel rate $C$ (bps)},
    ylabel style={at={(axis description cs:-0.1,0.5)}},
    ylabel={Mean delay (s)},
    legend style={at={(0.98,0.98)}, anchor=north east, font=\scriptsize, legend columns=4, rounded corners=2pt},
    legend cell align={left},
    every legend image post/.append style={xshift=-3pt},
    legend image code/.code={\draw[mark repeat=1, mark phase=1, #1] plot coordinates {(0cm,0cm)(0.2cm,0cm)};},
    grid=major,
    mark size=1.5pt
]

\draw[gray, dashed, line width=1pt] (axis cs:32.80,0.01) -- (axis cs:32.80,1000000);

\addplot[color=black!60!white, mark=*, line width=1.5pt] coordinates {
    (29.52, 145.860) (31.16, 108.125) (32.14, 87.346) (32.80, 74.189)
    (33.45, 61.551) (34.44, 43.501) (36.08, 17.528) (39.36, 1.128)
    (49.19, 0.191) (65.59, 0.102) (98.39, 0.063) (163.98, 0.037) (327.96, 0.019)
};

\addplot[color=red!80!black, mark=*, line width=1.5pt] coordinates {
    (29.52, 148.378) (31.16, 110.569) (32.14, 89.747) (32.80, 76.563)
    (33.45, 63.898) (34.44, 45.809) (36.08, 19.770) (39.36, 3.438)
    (49.19, 2.541) (65.59, 2.457) (98.39, 2.417) (163.98, 2.392) (327.96, 2.374)
};

\addplot[color=blue!60!white, mark=*, line width=1.5pt] coordinates {
    (29.52, 117.322) (31.16, 81.109) (32.14, 61.166) (32.80, 48.539)
    (33.45, 36.418) (34.44, 20.471) (36.08, 2.970) (39.36, 0.649)
    (49.19, 0.167) (65.59, 0.098) (98.39, 0.061) (163.98, 0.037) (327.96, 0.018)
};

\addplot[color=violet!20!white, mark=square*, line width=1.5pt] coordinates {
    (29.52, 387.979) (31.16, 331.877) (32.14, 301.086) (32.80, 281.602)
    (33.45, 262.887) (34.44, 236.154) (36.08, 194.849) (39.36, 128.466)
    (49.19, 5.163) (65.59, 3.383) (98.39, 2.720) (163.98, 2.191) (327.96, 1.794)
};

\addplot[color=violet!40!white, mark=square*, line width=1.5pt] coordinates {
    (29.52, 721.610) (31.16, 634.670) (32.14, 586.764) (32.80, 556.424)
    (33.45, 527.273) (34.44, 485.630) (36.08, 421.272) (39.36, 311.067)
    (49.19, 113.805) (65.59, 1.991) (98.39, 1.479) (163.98, 1.148) (327.96, 0.900)
};

\addplot[color=violet!60!white, mark=square*, line width=1.5pt] coordinates {
    (29.52, 1409.640) (31.16, 1286.448) (32.14, 1218.567) (32.80, 1175.576)
    (33.45, 1134.271) (34.44, 1075.263) (36.08, 984.070) (39.36, 824.481)
    (49.19, 473.385) (65.59, 138.662) (98.39, 0.864) (163.98, 0.627) (327.96, 0.453)
};

\addplot[color=violet!80!white, mark=square*, line width=1.5pt] coordinates {
    (29.52, 5550.775) (31.16, 5209.600) (32.14, 5021.605) (32.80, 4902.542)
    (33.45, 4788.148) (34.44, 4624.728) (36.08, 4372.170) (39.36, 3930.192)
    (49.19, 2957.843) (65.59, 1985.493) (98.39, 1013.144) (163.98, 235.264) (327.96, 0.118)
};

\addplot[color=green!60!black, mark=triangle*, line width=1.5pt] coordinates {
    (29.52, 10500.163) (31.16, 9898.494) (32.14, 9566.962) (32.80, 9356.991)
    (33.45, 9155.255) (34.44, 8867.061) (36.08, 8421.669) (39.36, 7642.234)
    (49.19, 5927.476) (65.59, 4212.718) (98.39, 2497.960) (163.98, 1126.154) (327.96, 97.324)
};
\legend{Shannon, AC, Huffman, rANS-K16, rANS-K8, rANS-K4, rANS-K1, Gzip}
\end{axis}
\end{tikzpicture}

%% file: figures/delay_vs_rate_moby_dick_llama.tex
\begin{tikzpicture}
\begin{axis}[
    width=\columnwidth,
    height=6cm,
    xmode=log,
    ymode=log,
    xmin=15,
    xmax=350,
    ymin=0.01,
    ymax=1000000,
    xlabel={Channel rate $C$ (bps)},
    ylabel style={at={(axis description cs:-0.1,0.5)}},
    ylabel={Mean delay (s)},
    legend style={at={(0.98,0.98)}, anchor=north east, font=\scriptsize, legend columns=4, rounded corners=2pt},
    legend cell align={left},
    every legend image post/.append style={xshift=-3pt},
    legend image code/.code={\draw[mark repeat=1, mark phase=1, #1] plot coordinates {(0cm,0cm)(0.2cm,0cm)};},
    grid=major,
    mark size=1.5pt
]

\draw[gray, dashed, line width=1pt] (axis cs:17.87,0.01) -- (axis cs:17.87,1000000);

\addplot[color=black!60!white, mark=*, line width=1.5pt] coordinates {
    (16.09, 182.536) (16.98, 133.575) (17.52, 106.598) (17.87, 89.512)
    (18.23, 73.110) (18.77, 49.678) (19.66, 22.441) (21.45, 2.929)
    (26.81, 0.375) (35.75, 0.139) (53.62, 0.071) (89.37, 0.040) (178.74, 0.020)
};

\addplot[color=red!80!black, mark=*, line width=1.5pt] coordinates {
    (16.09, 187.047) (16.98, 137.897) (17.52, 110.815) (17.87, 93.663)
    (18.23, 77.184) (18.77, 53.642) (19.66, 26.553) (21.45, 7.427)
    (26.81, 5.010) (35.75, 4.778) (53.62, 4.696) (89.37, 4.653) (178.74, 4.626)
};

\addplot[color=blue!60!white, mark=*, line width=1.5pt] coordinates {
    (16.09, 228.814) (16.98, 177.445) (17.52, 149.142) (17.87, 131.218)
    (18.23, 113.997) (18.77, 89.397) (19.66, 51.378) (21.45, 7.482)
    (26.81, 0.412) (35.75, 0.147) (53.62, 0.076) (89.37, 0.043) (178.74, 0.021)
};

\addplot[color=violet!20!white, mark=square*, line width=1.5pt] coordinates {
    (16.09, 737.341) (16.98, 646.526) (17.52, 596.486) (17.87, 564.793)
    (18.23, 534.343) (18.77, 490.844) (19.66, 423.618) (21.45, 312.324)
    (26.81, 100.689) (35.75, 4.134) (53.62, 3.151) (89.37, 2.485) (178.74, 1.986)
};

\addplot[color=violet!40!white, mark=square*, line width=1.5pt] coordinates {
    (16.09, 1366.566) (16.98, 1242.551) (17.52, 1174.216) (17.87, 1130.937)
    (18.23, 1089.355) (18.77, 1029.953) (19.66, 938.150) (21.45, 777.493)
    (26.81, 424.050) (35.75, 101.654) (53.62, 1.839) (89.37, 1.375) (178.74, 1.034)
};

\addplot[color=violet!60!white, mark=square*, line width=1.5pt] coordinates {
    (16.09, 2629.035) (16.98, 2438.532) (17.52, 2333.561) (17.87, 2267.079)
    (18.23, 2203.205) (18.77, 2111.956) (19.66, 1970.934) (21.45, 1724.147)
    (26.81, 1181.214) (35.75, 638.281) (53.62, 103.969) (89.37, 0.821) (178.74, 0.559)
};

\addplot[color=violet!80!white, mark=square*, line width=1.5pt] coordinates {
    (16.09, 10243.626) (16.98, 9652.324) (17.52, 9326.504) (17.87, 9120.152)
    (18.23, 8921.892) (18.77, 8638.663) (19.66, 8200.946) (21.45, 7434.942)
    (26.81, 5749.731) (35.75, 4064.520) (53.62, 2379.310) (89.37, 1031.141) (178.74, 23.908)
};

\addplot[color=green!60!black, mark=triangle*, line width=1.5pt] coordinates {
    (16.09, 20439.090) (16.98, 19311.184) (17.52, 18689.686) (17.87, 18296.070)
    (18.23, 17917.890) (18.77, 17377.632) (19.66, 16542.690) (21.45, 15081.539)
    (26.81, 11867.009) (35.75, 8652.479) (53.62, 5437.949) (89.37, 2866.325) (178.74, 937.607)
};
\legend{Shannon, AC, Huffman, rANS-K16, rANS-K8, rANS-K4, rANS-K1, Gzip}
\end{axis}
\end{tikzpicture}

%% file: sections/conclusion.tex
\section{Conclusion}
\label{sec:conclusion}

This paper has studied the compression--delay tradeoff that arises when a causal language model serves as the predictor within a predict-then-code architecture for real-time text transmission.
A queueing model was developed in which variable-length codewords are streamed over a constant-rate channel, and six coding schemes were compared under two LLM predictors.

A central finding is that modern tokenizers and LLM predictors have fundamentally reshaped the coder landscape.
By grouping characters into tokens of three to four characters on average, BPE and SentencePiece provide a natural coding unit over which Huffman's worst-case 1-bit overhead becomes a small fraction of the conditional entropy.
Combined with a strong LLM predictor, Huffman coding achieves 10--15\% overhead relative to Shannon with zero algorithmic delay, making it difficult to justify the deferred decodability of arithmetic coding or the block buffering of rANS in most practical regimes.
Arithmetic coding remains near-optimal in compression, but its intrinsic delay is only worthwhile near the capacity boundary where every fraction of a bit matters.
The remaining coders are less compelling: rANS's grouping advantage is largely subsumed by tokenization, and gzip cannot exploit the shared LLM predictor.

Perhaps more importantly, the experiments reveal that model quality shifts the operating point.
Llama~3.2's significant reduction in bits per character over GPT-2 can move the system from near-capacity to over-provisioned operation at the same channel rate, dramatically reducing queueing delay and changing which coder is optimal.
This effect is amplified by the practical reality of human speech: because speakers vary in rate, a communication system must provision the channel for the fastest speakers it serves, inherently over-provisioning for the majority.
In such a regime, Huffman coding paired with an LLM predictor is the pragmatic choice.

These observations shift the fundamental design tradeoff from an algorithmic question (which coder to deploy) to an AI engineering challenge (which predictor to deploy, at what inference cost).
As compute grows cheaper, larger models become increasingly viable, but there is also a practical middle ground: lightweight models obtained through distillation, quantization, or domain-specific adaptation may achieve sharp conditional distributions at inference speeds compatible with real-time speech.

Several directions remain open.
Sequential lossy compression under delay constraints could further reduce source rate and variability when controlled distortion is acceptable.
Real-time inference benchmarks on GPU and NPU hardware would quantify computational delay alongside the algorithmic delay studied here.
Validation on real wireless channels with error correction would bridge the gap between the idealized constant-rate channel model and practical deployments.
Finally, extending the predict-then-code framework to multimodal streaming sources, such as video or sensor data, would test whether the insights developed here for language generalize to higher-dimensional modalities.